# High-precision gigahertz-to-terahertz spectroscopy of aqueous salt solutions as a probe of the femtosecond-to-picosecond dynamics of liquid water

N. Q. Vinh[1,2,4*], Mark S. Sherwin[1,2], S. James Allen[1,2], D. K. George[4], A. J. Rahmani[4] and Kevin. W. Plaxco[1,3]

[1]Institute for Terahertz Science and Technology, University of California, Santa Barbara Santa Barbara, CA 93106, USA

[2]Department of Physics, University of California, Santa Barbara, Santa Barbara, CA 93106, USA

[3]Department of Chemistry and Biochemistry, University of California, Santa Barbara, Santa Barbara, CA 93106, USA

[4]Department of Physics, Virginia Tech, Blacksburg, VA 24061, USA

*corresponding author: vinh@vt.edu

**Running title: The picosecond dynamics of liquid water**

**Abstract**. Because it is sensitive to fluctuations occurring over femtoseconds to picoseconds, gigahertz-to-terahertz dielectric relaxation spectroscopy can provide a valuable window into water's most rapid intermolecular motions. In response, we have built a vector network analyzer dielectric spectrometer capable of measuring absorbance and index of refraction in this frequency regime with unprecedented precision. Using this to determine the complex dielectric response of water and aqueous salt solutions from 5.9 GHz to 1.12 THz (which we provide in the SI), we have obtained strong new constraints on theories of water's collective dynamics. For example, while the salt-dependencies we observe for water's two slower relaxations (8 and 1 ps) are easily reconciled with suggestions that they arise due to rotations of fully and partially hydrogen bonded molecules, respectively, the salt-dependence of the fastest relaxation (180 fs) appears difficult to reconcile with its prior assignment to liberations of single hydrogen bonds.

**INTRODUCTION**

Water is undoubtedly the most widely studied chemical system, with thousands of experimental and theoretical characterizations reported to date[1-12]. Detailed understanding of its collective molecular dynamics nevertheless remains an important outstanding problem in the physical and biological sciences[13-16]. The experimental literature describing water's femto-to-picosecond dielectric relaxations, for example, are often contradictory, with estimates for the timescale of the more rapid of these motions tending to cluster around the mutually exclusive values of 100 fs and 1 ps (table 1). This poor agreement, in turn, renders it difficult to distinguish between competing theoretical models of water's dynamical behavior, limiting our understanding of key aspects of this important and ubiquitous molecule[5, 13, 14, 17-22].

Because it can be used to monitor dipole fluctuations occurring over femtoseconds to picoseconds, gigahertz-to-terahertz dielectric relaxation spectroscopy could provide a valuable experimental window into the molecular-scale dynamics of water and aqueous solutions. Unlike infrared and Raman spectroscopy, for example, which are sensitive to the femtosecond-scale int*ra*molecular dynamics (i.e., bond vibrations), gigahertz-to-terahertz regime is sensitive to their picosecond-scale int*er*molecular dynamics (i.e., molecular motions involving hydrogen bond breaking). Unfortunately, however, daunting technical limitations, including the extremely strong absorbance of polar liquids and the often severe interference artifacts that occur at these wavelengths, have reduced the precision of prior spectroscopic studies (particularly at the upper end of this frequency range), limiting our ability to characterize any but the slowest and most strongly absorbing of water's dynamic modes. As recently as 2011, for example, Zasetsky[17] has stated that the "available experimental data in the wave number range of 1 to 100 cm$^{-1}$ [30 GHz to 3 THz] are limited, [and] the accuracy of these data is insufficient for an accurate characterization of fast processes."

In response to these observations we have built a vector network analyzer-based dielectric spectrometer capable of measuring the absorbance and index of refraction of water and aqueous solutions over the frequency range 5.9 GHz to 1.12 THz (0.27 to 51 mm or 0.2 to 37.4 cm$^{-1}$) with unprecedented precision, resolution and dynamic range[9]. This spectrometer, which is based on a vector network analyzer and frequency extension modules built by Virginia Diode Inc., achieves sub-100 Hz spectral resolution and a detection dynamic range spanning more than 15 orders of magnitude. Integrating the spectrometer with an automated, variable path length cell, which allows for high-precision measurement of how transmitted intensity and phase change with changing path length, we achieve precision of better than 0.2% in the measurement of both the real and imaginary components $\mathcal{E}(\nu)$ and $\mathcal{E}'(\nu)$ of the dielectric relaxation of water and aqueous solutions over the frequency range most relevant to its molecular-scale motions. The precision, spectral resolution and broad frequency range of these measurements[9] render them a unique asset with which to test theoretical models of water's dynamics over femotosecond to picosecond time scales.

**Table 1.** Relaxation times extracted from previous measurements of the dielectric function of water at 25˚C.

| Technique | Frequency range (GHz) | Slowest process, $\tau_D$ (ps)[a] | Next faster process, $\tau_2$ (ps)[a] | Fastest process, $\tau_3$ (ps)[a] | Refs |
|---|---|---|---|---|---|
| Dielectric relaxation | 68 – 75 | 8.17±0.36 | | | 23 |
| " | 1.1 – 57 | 8.27±0.02 | | | 7, 24-26 |
| " | Few data points up to 1000 | 8.24±0.02 | 0.21 | | 27 |
| " | 0.95-89 | 8.32±0.17 | 1.02±0.05 | | 28, 29 |
| " | 0.2-410 | 8.38±0.17 | 1.10±3.40 | | 30, 31 |
| " | 0.2-410 | 8.32±0.17 | 0.39-0.42 | | 13, 32-34 |
| " | 0.2-89 | 8.38 | 0.3 | | 35 |
| " | 600-3000 | | | 0.053 | 36 |
| Terahertz time-domain spectroscopy | 370-1620 | ~8 ps | ~0.25 | | 21 |
| " | 200-3500 | 8.3 | 0.25±0.01 | | 37 |
| " | 60-1500 | 8.24±0.40 | 0.18±0.14 | | 15, 38 |
| " | 100-2000 | 7.10±0.48 | 0.20±0.06 | | 14, 39, 40 |
| " | 200-1800 | | 0.336 | | 41 |
| Data collated from multiple experiments | Up to 25000 | 8.58 | 0.95 | 0.082 | 18 |
| " | Up to 1000 | 8.26±0.03 | 1.05±0.52 | 0.135±0.035 | 19 |
| " | Up to 7000 | 8.31 | 1.0 | 0.10 | 42 |
| **This work** | **5.9-1120** | **8.37±0.25** | **1.05±0.15** | **0.178±0.050** | |

[a]Error bars given are reported confidence intervals as provided in the literature. When no

values are reported, confidence intervals were lacking in the original literature.

**EXPERIMENTAL RESULTS**

We have measured the absorbance and phase shift of water and aqueous salt solutions over the three order of magnitude range 5.9 GHz to 1.12 THz as functions of path length, $d$, at 25.00 (±0.02)°C. In brief, our experimental set up[9, 15] consists of a variable-thickness, temperature controlled sample cell with which we can accurately measure *changes* in transmitted power as a function of *changes* in path length. Fitting this data to Beer's law, $I(d) = I_0\exp(-\alpha_{sol}d)$, with $I_0$ corresponding to the incident intensity, provides a precise (standard error of the mean of replicate measurements typically < 0.2%) measurement of the absorption coefficient of a sample, $\alpha_{sol}$ (Fig. 1), without the need for precise knowledge of the sample's absolute absorbance or absolute path length. In parallel we also fit the observed phase shift as a linear function of path length to define $n_{sol}$, the refractive index of the sample (Fig. 1, inset). Both properties of these samples are strong functions of frequency, monotonically increasing and decreasing, respectively, with rising frequency over this entire spectral range.

Knowledge of the absorbance and refractive index of our sample allow us to compute its complex dielectric response[9]:

$$\varepsilon'_{sol}(\nu) = n^2(\nu) - \kappa^2(\nu) = n^2(\nu) - (c\alpha(\nu)/4\pi\nu)^2$$
$$\varepsilon''_{sol}(\nu) = 2n(\nu) \cdot \kappa(\nu) \tag{1}$$

where $\nu$ is frequency, $n$ is the refractive index of the solution, and $\kappa$ is the extinction coefficient of the solution, which is related to the absorption coefficient, $\alpha$, by $\kappa = c\alpha/(4\pi\nu)$ with $c$ being the speed of light. From this we can calculate the complex dielectric function, $\varepsilon^*_{sol}$, of the solution, which, in turn provides a complete description of the interaction of the solute with the incoming electromagnetic wave.

The complex dielectric function $\varepsilon^*(\nu) = \varepsilon'(\nu) - i\varepsilon''(\nu)$, of pure polar liquids can be modeled as a sum of $n$ Debye relaxation terms[43]:

$$\varepsilon^{*} = \varepsilon_{\infty} + \frac{(\varepsilon_{s} - \varepsilon_{1})}{1 + i2\pi\nu\tau_{1}} + \frac{(\varepsilon_{1} - \varepsilon_{2})}{1 + i2\pi\nu\tau_{2}} + ... + \frac{(\varepsilon_{n-1} - \varepsilon_{\infty})}{1 + 2i\pi\nu\tau_{n}}$$
$$= \varepsilon_{\infty} + \sum_{j=1}^{n} \frac{\varepsilon_{j-1} - \varepsilon_{j}}{1 + i2\pi\nu\tau_{j}} = \varepsilon_{\infty} + \sum_{j=1}^{n} \frac{\Delta\varepsilon_{j}}{1 + i2\pi\nu\tau_{j}} \quad (2)$$

where $\varepsilon_{j-1}$ and $\varepsilon_{j}$ are the values of the dielectric constant at the beginning and at the end of the $j^{th}$ dispersion region, respectively; $\tau_{j}$ is the relaxation time of the $j^{th}$ Debye process and $\nu$ is the frequency of the incoming radiation. For consistency with the nomenclature in the prior literature we use here $\tau_{D}$ to denote $\tau_{1}$ (the first Debye process). $\Delta\varepsilon_{j} = \varepsilon_{j-1} - \varepsilon_{j}$ is the weighted contribution of each Debye process to the total relaxation. With $j=n$, $\varepsilon_{j=n} = \varepsilon_{\infty}$ captures the contributions to the dielectric function from modes at frequencies much greater than our experimental range (including previously-reported modes at 5 and 15 THz; ref[42]) and thus reflects contributions from molecular oscillation dynamics. With $j=1$, $\varepsilon_{j-1=0}$ is equivalent to $\varepsilon_{s}$, the static (or direct current) dielectric constant, which we have assigned a value of 78.38 for pure water at 25°C (from previous studies)[6, 30, 44]. Given this, the real (or dispersive) $\varepsilon'(\nu)$ and imaginary (or lossy) $\varepsilon''(\nu)$ parts of the complex permittivity can be derived from eq. 2:

$$\varepsilon'(\nu) = \varepsilon_{\infty} + \sum_{j=1}^{n} \frac{\Delta\varepsilon_{j}}{1 + (2\pi\nu\tau_{j})^{2}}$$
$$\varepsilon''(\nu) = \sum_{j=1}^{n} \frac{\Delta\varepsilon_{j} 2\pi\nu\tau_{j}}{1 + (2\pi\nu\tau_{j})^{2}} \quad (3)$$

Using this approach we have determined the real and imaginary parts of the complex permittivity of liquid water over the gigahertz to terahertz regime with unprecedented resolution and precision (Fig. 2). As a resource for the community of researchers investigating the intermolecular dynamics of water, we provide these novel data in the S.I.

We believe the high precision data we present here will prove an important tool with which to test models of the femto-to-picosecond dynamics of liquid water. For example, although previous single-experiment studies of these dynamics have almost invariably been interpreted in terms of the existence of two dielectric relaxations, the large discrepancy between the purported time constants of the more rapid of these two phases (table 1) led Ellison[18], Beneduci[19], and Vij et al.[42] to independently conclude that the dynamics of water over this timescale are more accurately described as the sum of three Debye relaxations. To test this hypothesis, however, these prior authors combined data from multiple, disperate experimental approaches. The data set we have obtained, in contrast, is of sufficient quality to confirm this hypothesis with good statistical significance using a single, internally consistent data set. To see this we fit the real and imaginary components of the complex permittivity we observed for pure water using a model containing either two or three Debye relaxation processes ($n = 2$ or 3 in eqs. 2 and 3). [Note: we performed the fit as a fit of $\log(\varepsilon^*)$ versus the logarithm of the dielectric response (i.e., we took the logarithm of both sides of eq. 2) in order to suppress the otherwise overwhelming contribution of the slowest relaxation to the global fit; the amplitude of term in eq. 2 associated with the slowest relaxation is two orders of magnitude larger than those associated with the other, faster relaxations, and thus it is so heavily weighted in a linear fit that the precision with which we can determine the parameters associated with the (much smaller) faster relaxations is reduced. This said, fits to eq. 2 (i.e., in its original, linear form) produce time constants and amplitudes within uncertainty of those seen for the logarithmic fit, albeit with larger confidence intervals (see table SI1[45]).]

In our fits the four parameters in the two-Debye model, $\varepsilon_1$, $\varepsilon_\infty$, $\tau_D$ and $\tau_2$, are varied simultaneously and the static permittivity, $\varepsilon_s$, is held fixed at the literature value[13, 18, 24, 44]. The best-fit values of $\varepsilon_1$ and $\varepsilon_\infty$ are $5.48 \pm 0.10$ and $3.5 \pm 0.3$, respectively; the latter appears to be somewhat higher than the value of 3.06 (no confidence intervals reported) previously derived[46] from independent observations of refractive index of pure water at 4.5 THz. The best-fit values of the two time constants are $\tau_D = 8.23 \pm 0.25$ ps and $\tau_2 = 410 \pm 150$ fs (Fig. 2, top; adjusted $R^2 = 0.99979$; reduced $\chi^2 = 3.65 \times 10^{-5}$). The former agrees closely with previously reported values (table 1). The time constant we obtain for the latter, more rapid process, however, differs significantly from previously reported values, which tend to cluster around either 1 ps or 100 fs (table 1).

More detailed inspection of our data suggests that the latter discrepancy occurs because, as suspected, the two-Debye model is insufficient to explain the relaxation dynamics of water[18,19,42]. To show this we fit our data to two more complex models. The first is the three-Debye model suggested earlier by Vij et al.[42], Ellison[18], and Beneduci[19]. The second is a two-Debye-plus-resonance model likewise promulgated by numerous previous authors[13,41,42].

The three-Debye model, in which we vary $\tau_D$, $\tau_2$, $\tau_3$, $\varepsilon_1$, $\varepsilon_2$ and $\varepsilon_\infty$, achieves a significantly improved correlation coefficient relative to the two-Debye model (Fig. 2, top; adjusted $R^2 = 0.99997$; reduced $\chi^2 = 4.65 \times 10^{-6}$), significantly reduced fit residuals (root-mean-squared residuals: 0.002 versus 0.007 for the real components and 0.002 versus 0.005 for the imaginary components), and significantly reduced serial correlations of the residuals ($R^2$ of 0.954 versus 0.995 for the real and 0.945 versus 0.992 for the imaginary; see Fig. SI2[45] for comparison of the residuals). The fit to the three-Debye model produces $\varepsilon_\infty = 3.2 \pm 0.3$, which is now within experimental uncertainty of the prior literature value[46] and $\varepsilon_1$, $\varepsilon_2$ of $6.05 \pm 0.10$ and $4.91 \pm 0.10$, respectively. Not surprisingly, this three Debye model recovers the same large amplitude, slower relaxation apparent in previous dielectric relaxation spectroscopy studies ($\tau_D = 8.37 \pm 0.25$ ps, relative amplitude $96.2 \pm 0.3\%$). In addition, the three Debye model outputs relaxations with time constants that, at $1.05 \pm 0.15$ ps and $178 \pm 50$ fs (relative amplitudes of $1.5 \pm 0.3\%$ and $2.3 \pm 0.3\%$ respectively), correspond closely to *each* of the two faster processes so often reported in the prior literature (table 1).

In addition to the two- and three-Debye models we have, for completeness, also fitted our data using a model that includes a resonant process suggested by prior authors as an important component in the femto-to-picosecond dynamics of liquid water[13, 41, 42]. Specifically, we have added a damped resonance term to eq. 2.

$$\varepsilon^* = \varepsilon_\infty + \sum_{j=1}^{n} \frac{(\varepsilon_{j-1} - \varepsilon_j)}{1 + i 2\pi\nu\tau_j} + \frac{A_{OSC}/(2\pi)^2}{\nu_{OSC}^2 - \nu^2 + i\nu(k_{OSC}/2\pi)} \quad (4)$$

where $A_{OSC}/(2\pi)^2$, $\nu_{OSC}$, and $k_{OSC}$ are, respectively, the spectral amplitude of the damped resonance, its frequency, and its damping coefficient. We fit our data to a version of this model that includes two Debye relaxations (i.e., $n = 2$) and one damped resonance, varying the parameters $\varepsilon_1$, $\varepsilon_\infty$, $\tau_D$, $\tau_2$, $\nu_{OSC}$, $A_{OSC}$ and $k_{OSC}$ but, as for the two- and three-Debye models described above, fixing the static permittivity, $\varepsilon_s = \varepsilon_{j-1=0}$, to its literature value[6, 18, 30, 44]. Doing so we obtain best-fit values of $\varepsilon_1 = 5.23 \pm 0.10$, $\varepsilon_\infty = 3.52 \pm 0.30$, $\tau_D = 8.36 \pm 0.25$ ps, $\tau_2 = 0.45 \pm 0.15$ ps, $\nu_{osc} = 1.4 \pm 0.1$ THz, $A_{osc}/(2\pi)^2 = 39.9 \pm 0.5$ THz$^2$, and $k_{osc} = 7.4 \pm 0.5$ THz. However, while this fit is reasonably good, the best-fit value of $\tau_2$ differs rather significantly with estimates from the prior literature (table 1). Moreover, the fit statistics for the two-Debye-plus-damped-resonance model are poorer than those of the three Debye model (adjusted $R^2 = 0.99993$; reduced $\chi^2 = 1.3 \times 10^{-5}$ for the two-Debye-plus-resonance model versus $R^2 = 0.99997$ and reduced $\chi^2 = 4.65 \times 10^{-6}$ for the three-Debye model). And despite the fact that the two-Debye-plus-resonance model *contains one more free parameter* than the three-Debye model, both the fit residuals (root-mean-squared residuals: 0.002 versus 0.004 for both the real and imaginary components) and their serial correlations ($R^2$ of 0.954 versus 0.980 for the real and 0.945 versus 0.985 for the imaginary) are improved for the three-Debye model (see Fig. SI2[45] for comparisons of the residuals). Taken together, these observations suggest that the three Debye model provides a more accurate description of the femto-to-picosecond dynamics of liquid water.

Our ability to, for the first time, robustly and simultaneously quantify all three Debye processes in a single experimental data set stems from the improved signal-to-noise ratio of our spectrometer and the broad range of frequencies it spans. Specifically, the previously reported ~1 ps value is associated with frequency-domain spectroscopy conducted below 410 GHz. The lack of *high*-frequency coverage in these studies would render them relatively insensitive to low-amplitude relaxation process occurring at time scales faster than ~1 ps. The reported sub-picosecond time constants, in contrast, arise primarily from time-domain terahertz spectroscopic studies conducted over the range of 100 GHz to 2 THz. And thus the insufficient *low*-frequency coverage of these studies could render them insensitive to low amplitude processes that are slower than ~200 fs. The spectral range of our spectrometer, in contrast, spans all of the relevant time scales, thus largely alleviating these potentially important biases.

To provide further tests of theoretical models of the femto-to-picosecond dynamics of liquid water we have also characterized the dielectric relaxation responses of aqueous sodium chloride (NaCl) solutions, collecting data at concentrations as high as 3 M (corresponding to 18% w/v). For an electrolyte solution of conductivity $\sigma$, the dielectric properties of the solutions are in the form:

$$\varepsilon^* = \varepsilon_\infty + \Delta\varepsilon_u + \Delta\varepsilon_{sw} + \sum_{j=1}^{n} \frac{(\varepsilon_{j-1} - \varepsilon_j)}{1 + i2\pi\nu\tau_j} + i\frac{\sigma}{2\pi\nu\varepsilon_0} \quad (5)$$

where $\varepsilon_0$ is the permittivity of free space. The second and the third terms of eq. 5, $\varepsilon_u$, and $\varepsilon_{sw}$, respectively, reflect the contributions of the tumbling of the ion (and its associated hydration shell) and the rotational relaxation of water molecules within the hydration shell. However, due to, respectively, the increased mass and strong Coulombic field of solvated ions these terms do not contribute significantly to the relaxation dynamics of water over the frequency range under investigation[7, 21, 44, 47] and can be neglected in our analysis. The fourth term in this equation, reflects the contributions of the solvent (water) to the dielectric response of the solution and the final term reflects Ohmic losses due to the conductivity of electrolyte solutions.

Our high-resolution spectra of aqueous salt solutions provide a vehicle to further explore the relative merits of the two-Debye, three-Debye, and two-Debye-plus-resonance models. Specifically, we first fit these data to eq. 5 (with $n = 2$ or 3) varying the parameters $\varepsilon_j$, $\varepsilon_\infty$, $\tau_j$ and $\varepsilon_s$ but fixing the measured electrical conductivity, $\sigma$, using independently derived, salt-concentration-dependent valued derived from measurements of bulk conductivity (see Fig. SI1[45]). Once again we find that the three-Debye model provides a statistically significant improvement in the fits of both the real and the imaginary components relative to the two-Debye model. For example, for data collected at 2 M (Fig. 2, bottom) both the reduced $\chi^2$ and adjusted $R^2$ are improved (1.32 × 10$^{-5}$ versus 1.17 × 10$^{-4}$ and 0.9999 versus 0.9991) for the three-Debye model, as are the fit residuals (root-mean-squared residuals: 0.004 versus 0.011 for the real components and 0.003 versus 0.011 for the imaginary components; serial correlation of the residuals: $R^2 = 0.976$ versus 0.997 for the real components and $R^2 = 0.946$ versus 0.996 for the imaginary components). And, as was true for our studies of pure water, the best fit value of $\varepsilon_\infty$ for our salt solutions, 3.2 ± 0.3, is also in good agreement with previous observations[46]. We then fit our salt solution data to a two-Debye-plus-resonance model (eq. SI1[45]), but, as was true for the case with pure water, we once again do not find any significant improvement in the fit (Fig. SI2[45] and table SI2[45]). Furthermore, the best-fit resonance frequency $\nu_{OSC}$ shifts from 1.4 ± 0.1 THz for pure water to 1.1 ± 0.1 THz at 2 M salt, increasing the difficulty of assigning this putative resonance to a plausible physical mechanism. It thus appears that, as it does for pure water, the three Debye model accurately captures the femto-to-picosecond relaxational dynamics of water even in the presence of very high concentrations of dissolved salt.

In contrast to prior studies of aqueous salt solutions[41, 44, 47, 48], the time constants of the relaxations we observe are effectively *in*dependent of salt concentration over the entire concentration range we have explored (Fig. 3, top). Specifically, fitting our data to the three-Debye model (and fixing $\sigma$ as above using independently obtained measurements of direct current conductivity) we obtain relaxation time constants within experimental uncertainty of the mean values of 8.05 ± 0.55 ps, 1.25 ± 0.35 ps and 185 ± 50 fs across all salt concentrations. These values are likewise within uncertainty of the values obtained above for pure water. The amplitudes of the three phases, in contrast, vary strongly with increasing salt (Fig. 3, bottom). Specifically, while the amplitude of the slowest process decreases significantly (by 48% at 3 M) with increasing salt concentrations, those of the two more rapid processes increase (by 260% and 53%, respectively, at 3 M). The differential salt-dependence of these amplitudes would explain why, in contrast to our observations, prior authors[44, 47] found that the time constants of water's faster relaxation is also salt-dependent. Specifically, prior studies assumed that the relaxation of water is a two-Debye process. As the salt concentration rises and the relative amplitudes of the two more rapid phases shift, the weighted contributions of these two phases to the (assumed) single rapid process would also change, changing its best-fit time constant.

**DISCUSSION**

The theoretical literature regarding the dynamics of liquid water is voluminous; so much so that we cannot provide an exhaustive review of it much less an exhaustive discussion of how these theories fair in light of our new, high-resolution data set.

Nevertheless, we discuss here in brief some highlights illustrating the ways in which our data set can shed light on water's most rapid intermolecular motions.

Previous reports argue that the 8 ps process that dominates the dielectric relaxation of water, and that has likewise been observed by NMR[49-51] and polarization-selective pump-probe experiments[52-54], arises due to cooperative reorientation of the water dipole. Specifically, molecular dynamics studies ascribe this process to a Debye relaxation involving the re-orientation of water molecules engaged in their full tetrahedral complement of hydrogen bonds[5, 8, 10, 17, 20, 55-58]. Examples include the model of Laage et al.[5, 20], who describe the complete molecule reorientation via angular jumps and frame tumbling occurring at ~6 ps, and the model of Zasetzky[17] which consists of reorientations of fixed axis dipoles in the double well potential set up by the four neighboring waters such that from thermally activated escape from this potential well occurs with a time constant of ~5 ps. Because most of the molecules in liquid water participate in their full complement of hydrogen bonds[30], and because the relevant motion entails a large-scale reorientation of the water dipole, such processes would be expected to contribute significantly to water's overall dielectric relaxation response. This is consistent with our data suggesting that this process contributes 96.2% of the total dielectric response of pure water at these frequencies.

The salt-independence of the time constant of the slowest relaxation and the strong salt-dependence of its amplitude are consistent with the above mechanistic assignment. Specifically, previous researchers have argued that sodium and chloride ions do not affect the bulk structure of water[53, 59], suggesting that they would not alter the rate of this large-scale conformational rearrangement. In contrast to its rate, however, the amplitude of the 8 ps relaxation changes quite appreciably as a function of salt concentration. Specifically, its absolute amplitude decreases by 48% as the salt concentration is raised from zero to 3 M. The concentration of water in these solutions, in contrast, falls by only 6% (from 55.5 to 52.1 M) over this same range of salt concentrations. The discrepancy in these two numbers has previously been attributed[7, 47, 58] to two mechanisms, both of which presumably contribute. First, water molecules in the tightly-bound first hydration shell surrounding each ion likely do not participate in this slowest relaxation process due to the reduced orientational polarization of water in a strong Coulombic field of ions, an effect known as dielectric saturation. Second, the movement of ions in an electric field will reorient water molecules in opposition to the field due to kinetic depolarization, an effect that arises due to a coupling between the dielectric and hydrodynamic properties of liquids. Given the assumptions that any water molecules in the hydration shell do not participate in this slowest relaxational process, the reduced amplitude of this process at high salt provides a means of determining the number of water molecules per equivalent of electrolyte, that are unable to contribute to the solvent relaxation process. The value we obtain, ~4.8 ± 0.5 water molecules bound to each sodium ion (see methods in Supporting Information, and note that prior dielectric response studies have shown that the dynamics of water molecules in the hydration shell of chloride ion are virtually indistinguishable those of from bulk water[35, 60]), is in good agreement with estimates of the size of the solvation shell of this ion derived previously using continuum theory approaches[44, 47].

The next more rapid relaxation process that we observe, which is also observed in polarization-selective pump-probe experiments[52, 53], occurs at ~1 ps. Many molecular dynamics-based studies have suggested that this relaxation corresponds to the reorientation of "weakly-bound" water molecules lacking at least three of their four potential hydrogen bonds[8, 10, 26, 55-59]. This interpretation, however, is not universally accepted: Zasetsky, for example, has suggested that it arises due to "intrawell relaxation" of water molecules residing in the potential double well set up by its four closest neighbors[17]. Our data supports the former, "weakly-bound waters" assignment for this more rapid mode. For example, the 1.5% relative amplitude we observe for this intermediate time scale process is reasonably consistent with previous estimates that only 2-4% of the molecules in room temperature water participate in one or fewer hydrogen bonds[42]. Perhaps more compellingly, this mechanistic assignment is also supported by our observation that, while the rate of this process is independent of salt concentration, its absolute amplitude increases strongly with increasing salt. Specifically, its absolute amplitude increases by 260% as the salt concentration rises from 0 to 3 M despite the fact that the concentration of water in the solution *falls* by 6% (or more if, as appears to be true for the relaxation underlying the slowest phase, the water molecules in the hydration shell do not participate in this relaxation). We believe this occurs because, as has been suggested by molecular dynamics simulations[61], solute-driven disruption of water's precisely hydrogen-bonded structure leads to an increase in the number of partially hydrogen bonded waters with increasing solute concentration. In contrast, our observations are difficult to reconcile with Zasetsky's assignment of this mode to intrawell relaxation[17], as the number of water molecules residing in the appropriate potential double well (which requires four tetrahedrally arranged neighbors) would be expected to fall rather than to rise as the concentration of water falls with increasing salt.

The fastest of the three relaxations we observe, which exhibits a 180 fs time constant and 2.3% relative amplitude, is also observed via time-domain terahertz[14, 15, 21, 37-41] spectroscopy. Molecular dynamics and quantum mechanical computations suggest that this phase arises due to the rapid disruption and reformation of individual hydrogen bonds[5, 8, 10, 55-58]. That is, this process may reflect water molecules that liberate slightly from their most stable hydrogen bonding geometry, breaking a single hydrogen bond before relaxing back to the same position and reforming the bond. Consistent with this argument, the rate we observe for this process is independent of salt concentration, which would be expected for a process associated with the breaking and reforming of single hydrogen bonds. This widely cited mechanistic assignment is less easy to reconcile, however, with our observation that, like that of the intermediate relaxation, the absolute amplitude of this fastest relaxation increases (by 53% at 3 M) with rising salt concentrations. Specifically, given that the concentration of water molecules, and thus the total number of hydrogen bonds per unit volume, decreases with increasing salt[61] our observations appear inconsistent with the assignment of the fastest relaxation process to the breaking and reforming of single hydrogen bonds, the number of which would likewise be expected to decrease.

**ACKNOWLEDGEMENT**

This work was supported by a grant from the W. M. Keck foundation and the Institute of Critical Technology and Applied Sciences (ICTAS) at Virginia Tech.

**References:**

1. M. Chaplin, Nat Rev Mol Cell Bio **7**, 861-866 (2006).
2. Y. Levy and J. N. Onuchic, Annu Rev Bioph Biom **35**, 389-415 (2006).
3. P. Ball, Chemphyschem **9**, 2677-2685 (2008).
4. D. P. Zhong, S. K. Pal and A. H. Zewail, Chem Phys Lett **503**, 1-11 (2011).
5. D. Laage and J. T. Hynes, Science **311**, 832-835 (2006).
6. W. J. Ellison, K. Lamkaouchi and J. M. Moreau, J Mol Liq **68**, 171-279 (1996).
7. U. Kaatze, J Solution Chem **26**, 1049-1112 (1997).
8. I. Ohmine and H. Tanaka, Chem Rev **93**, 2545-2566 (1993).
9. N. Q. Vinh, S. J. Allen and K. W. Plaxco, J Am Chem Soc **133**, 8942-8947 (2011).
10. U. Kaatze, R. Behrends and R. Pottel, J Non-Cryst Solids **305**, 19-28 (2002).
11. A. Mandal, K. Ramasesha, L. De Marco and A. Tokmakoff, J Chem Phys **140** (20), 204508 (2014).
12. E. H. Backus, K. J. Tielrooij, M. Bonn and H. J. Bakker, Optics letters **39** (7), 1717-1720 (2014).
13. T. Fukasawa, T. Sato, J. Watanabe, Y. Hama, W. Kunz and R. Buchner, Phys Rev Lett **95**, 197802-197804 (2005).
14. C. Ronne, P. O. Astrand and S. R. Keiding, Phys Rev Lett **82**, 2888-2891 (1999).
15. J. T. Kindt and C. A. Schmuttenmaer, J Phys Chem **100** (24), 10373-10379 (1996).
16. E. W. Castner and M. Maroncelli, J Mol Liq **77**, 1-36 (1998).
17. A. Y. Zasetsky, Phys Rev Lett **107** (11) (2011).
18. W. J. Ellison, J Phys Chem Ref Data **36**, 1-18 (2007).
19. A. Beneduci, J Mol Liq **138**, 55-60 (2008).
20. D. Laage and J. T. Hynes, J Phys Chem B **112** (45), 14230-14242 (2008).
21. K. J. Tielrooij, N. Garcia-Araez, M. Bonn and H. J. Bakker, Science **328**, 1006-1009 (2010).
22. S. Funkner, G. Niehues, D. A. Schmidt, M. Heyden, G. Schwaab, K. M. Callahan, D. J. Tobias and M. Havenith, J Am Chem Soc **134** (2), 1030-1035 (2012).
23. K. E. Mattar and H. A. Buckmaster, J Phys D Appl Phys **23**, 1464-1467 (1990).
24. U. Kaatze, J Chem Eng Data **34**, 371-374 (1989).
25. U. Kaatze, Chem Phys Lett **203**, 1-4 (1993).
26. P. Petong, R. Pottel and U. Kaatze, J Phys Chem A **104**, 7420-7428 (2000).
27. H. J. Liebe, G. A. Hufford and T. Manabe, Int J Infrared Milli **12**, 659-675 (1991).
28. J. Barthel, K. Bachhuber, R. Buchner and H. Hetzenauer, Chem Phys Lett **165**, 369-373 (1990).
29. J. Barthel and R. Buchner, Pure Appl Chem **63**, 1473-1482 (1991).
30. R. Buchner, J. Barthel and J. Stauber, Chem Phys Lett **306**, 57-63 (1999).
31. J. B. Hasted, S. K. Husain, F. A. M. Frescura and J. R. Birch, Infrared Phys **27**, 11-15 (1987).
32. T. Sato and R. Buchner, J Phys Chem A **108**, 5007-5015 (2004).
33. T. Sato and R. Buchner, J Chem Phys **118**, 4606-4613 (2003).
34. J. B. Hasted, S. K. Husain, F. A. M. Frescura and J. R. Birch, Chem Phys Lett **118**, 622-625 (1985).
35. D. A. Turton, J. Hunger, G. Hefter, R. Buchner and K. Wynne, J Chem Phys **128** (16) (2008).
36. M. S. Zafar, J. B. Hasted and Chamberl.J, Nature-Phys Sci **243**, 106-109 (1973).

37. H. Yada, M. Nagai and K. Tanaka, Chem Phys Lett **464**, 166-170 (2008).
38. D. S. Venables and C. A. Schmuttenmaer, J Chem Phys **108**, 4935-4944 (1998).
39. C. Ronne, L. Thrane, P. O. Astrand, A. Wallqvist, K. V. Mikkelsen and S. R. Keiding, J Chem Phys **107**, 5319-5331 (1997).
40. C. Ronne and S. R. Keiding, J Mol Liq **101**, 199-218 (2002).
41. M. Kondoh, Y. Ohshima and M. Tsubouchi, Chem Phys Lett **591**, 317-322 (2014).
42. J. K. Vij, D. R. J. Simpson and O. E. Panarina, J Mol Liq **112**, 125-135 (2004).
43. P. Debye, (Chemical Catalog, New York, 1929).
44. R. Buchner, G. T. Hefter and P. M. May, J Phys Chem A **103**, 1-9 (1999).
45. See supplemental material at [URL will be inserted by AIP].
46. H. R. Zelsmann, J Mol Struct **350** (2), 95-114 (1995).
47. K. Nortemann, J. Hilland and U. Kaatze, J Phys Chem A **101**, 6864-6869 (1997).
48. Y. Hayashi, Y. Katsumoto, S. Omori, N. Kishii and A. Yasuda, J Phys Chem B **111** (5), 1076-1080 (2007).
49. D. W. G. Smith and J. G. Powles, Mol Phys **10**, 451-463 (1966).
50. E. Lang and H. D. Ludemann, J Chem Phys **67**, 718-723 (1977).
51. J. Jonas, T. Defried and D. J. Wilbur, J Chem Phys **65**, 582-588 (1976).
52. H. J. Bakker, S. Woutersen and H. K. Nienhuys, Chem Phys **258**, 233-245 (2000).
53. S. Woutersen, U. Emmerichs and H. J. Bakker, Science **278**, 658-660 (1997).
54. J. J. Loparo, C. J. Fecko, J. D. Eaves, S. T. Roberts and A. Tokmakoff, Phys Rev B **70**, 180201-180204 (R) (2004).
55. H. Tanaka and I. Ohmine, J Chem Phys **87**, 6128-6139 (1987).
56. D. Bertolini, M. Cassettari, M. Ferrario, G. Salvetti and P. Grigolini, Chem Phys Lett **98**, 548-553 (1983).
57. D. Bertolini, M. Cassettari, M. Ferrario, G. Salvetti and P. Grigolini, J Chem Phys **81**, 6214-6223 (1984).
58. I. Ohmine, H. Tanaka and P. G. Wolynes, J Chem Phys **89** (9), 5852-5860 (1988).
59. Y. J. Zhang and P. S. Cremer, Curr Opin Chem Biol **10**, 658-663 (2006).
60. W. Wachter, W. Kunz, R. Buchner and G. Hefter, J Phys Chem A **109** (39), 8675-8683 (2005).
61. A. Chandra, Phys Rev Lett **85** (4), 768-771 (2000).

**Figure caption:**

**Figure 1:** The gigahertz-to-terahertz absorption of both pure liquid water and aqueous salt solutions increase monotonically with rising frequency. (**inset**) The refractive indices of water and salt solutions, in contrast, decrease with increasing frequency. The data described here and in the following figures were collected at 25.0°C.

**Figure 2:** Gigahertz-to-terahertz dielectric response of (**top**) water and (**bottom**) 2 M NaCl in water provides insights into the liquid's dynamics over the femtosecond to picosecond timescale. The red curves are fits of the real and the imaginary components of the complex dielectric response of the two samples to a three-Debye model. A comparison of the fit residuals for the two- and three-Debye models (upper and lower **insets**, respectively) suggests that the relaxational dynamics of water and aqueous salt solutions are best described as the sum of three Debye processes. This is evidenced by, for example, the significantly reduced magnitude and serial correlation coefficients of the residuals obtained when a third Debye process is added to the model (see text for more detailed statistical analysis).

**Figure 3:** The extent to which water's molecular-scale relaxations vary with increasing salt concentration provides insights into their mechanistic origins. (**top**) The time constants of all three of water's femto-to-picosecond scale relaxations are effectively independent of salt concentration (the best-fit slopes are all within experimental uncertainty of zero). (**bottom**) Their amplitudes, in contrast, vary significantly with changing salt concentration: while the amplitude of the slowest relaxation falls with rising salt concentrations, the amplitudes of the two faster components increase. The mechanistic implications of these effects are explored in the text.

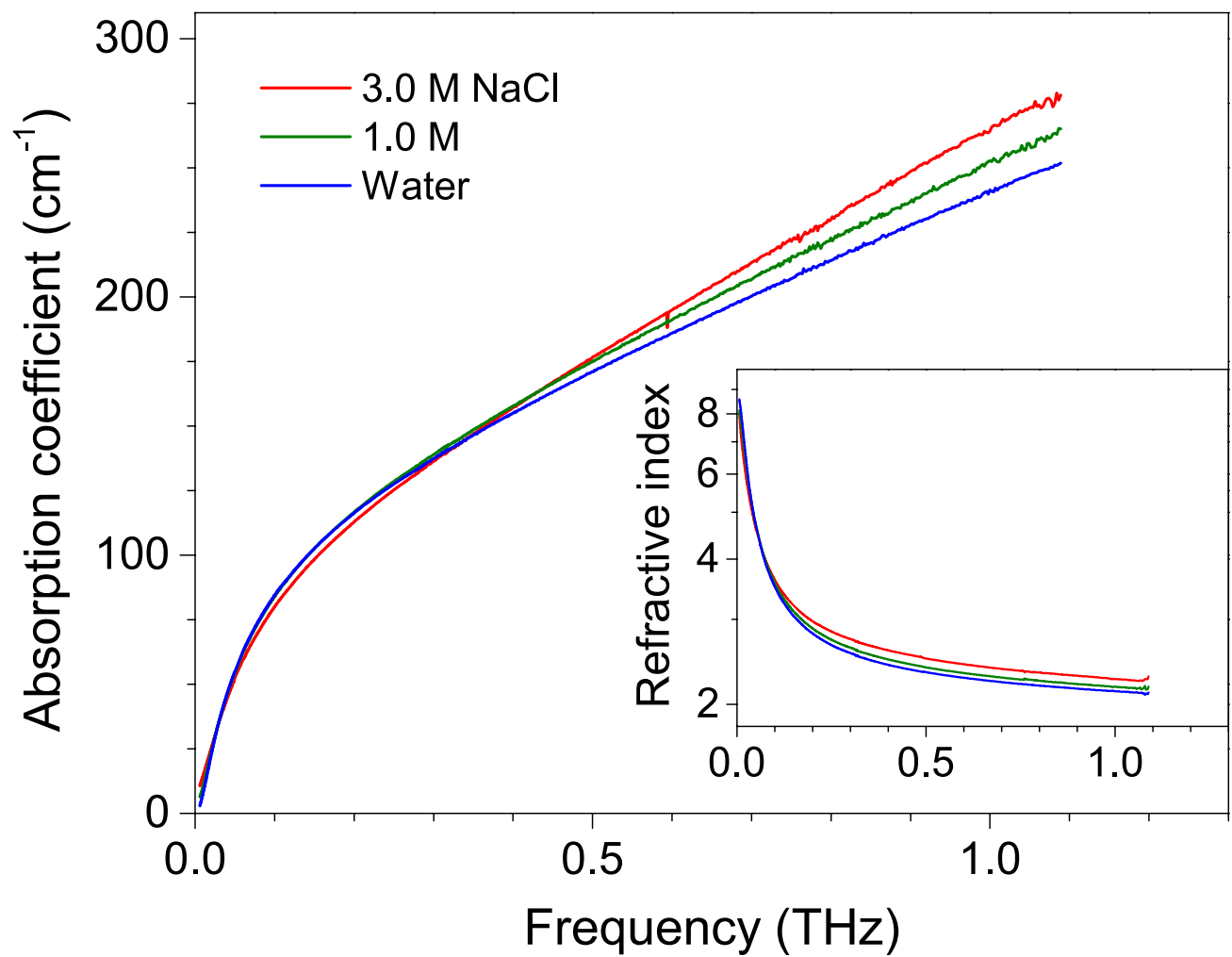

Figure 1

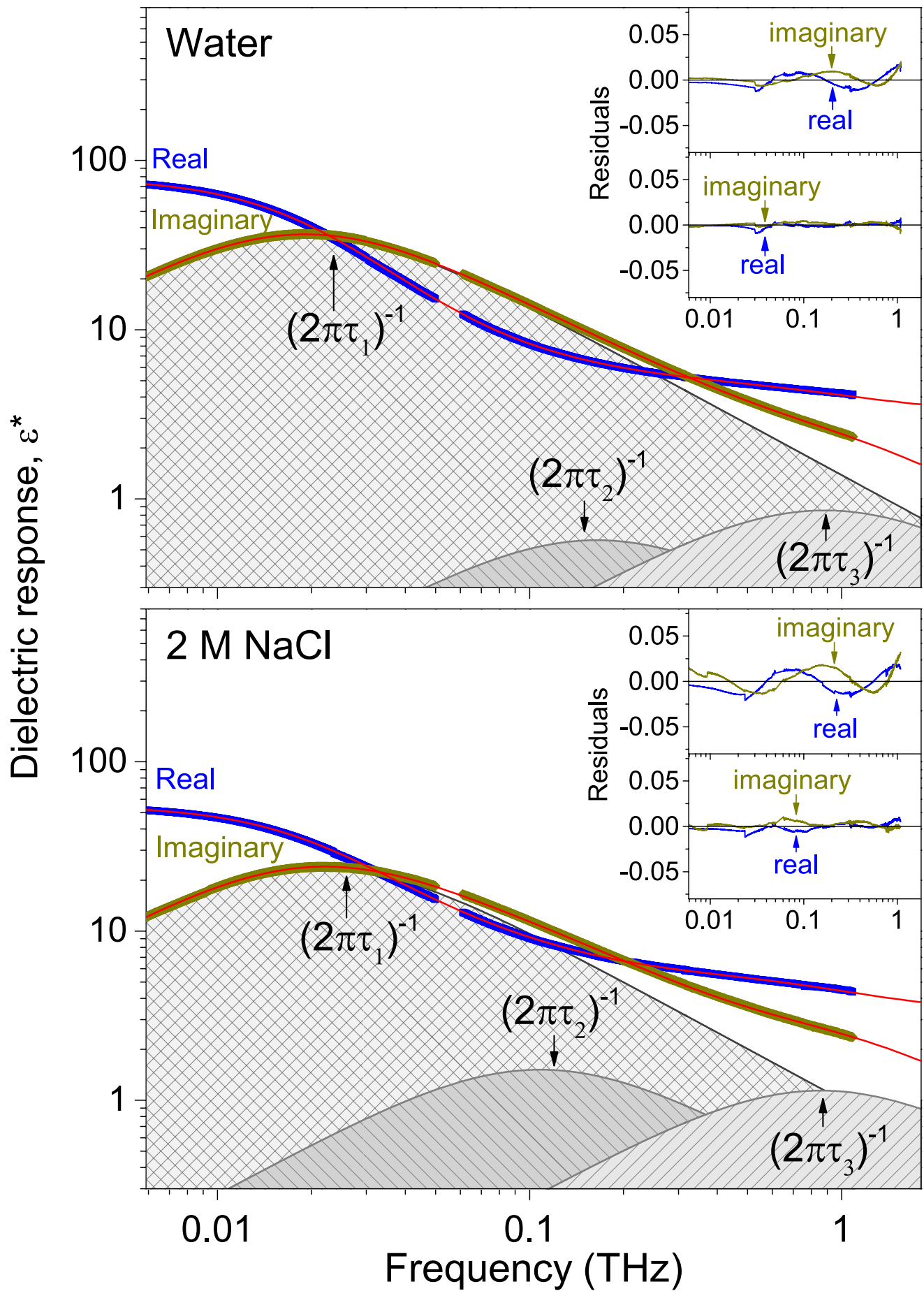

Figure 2

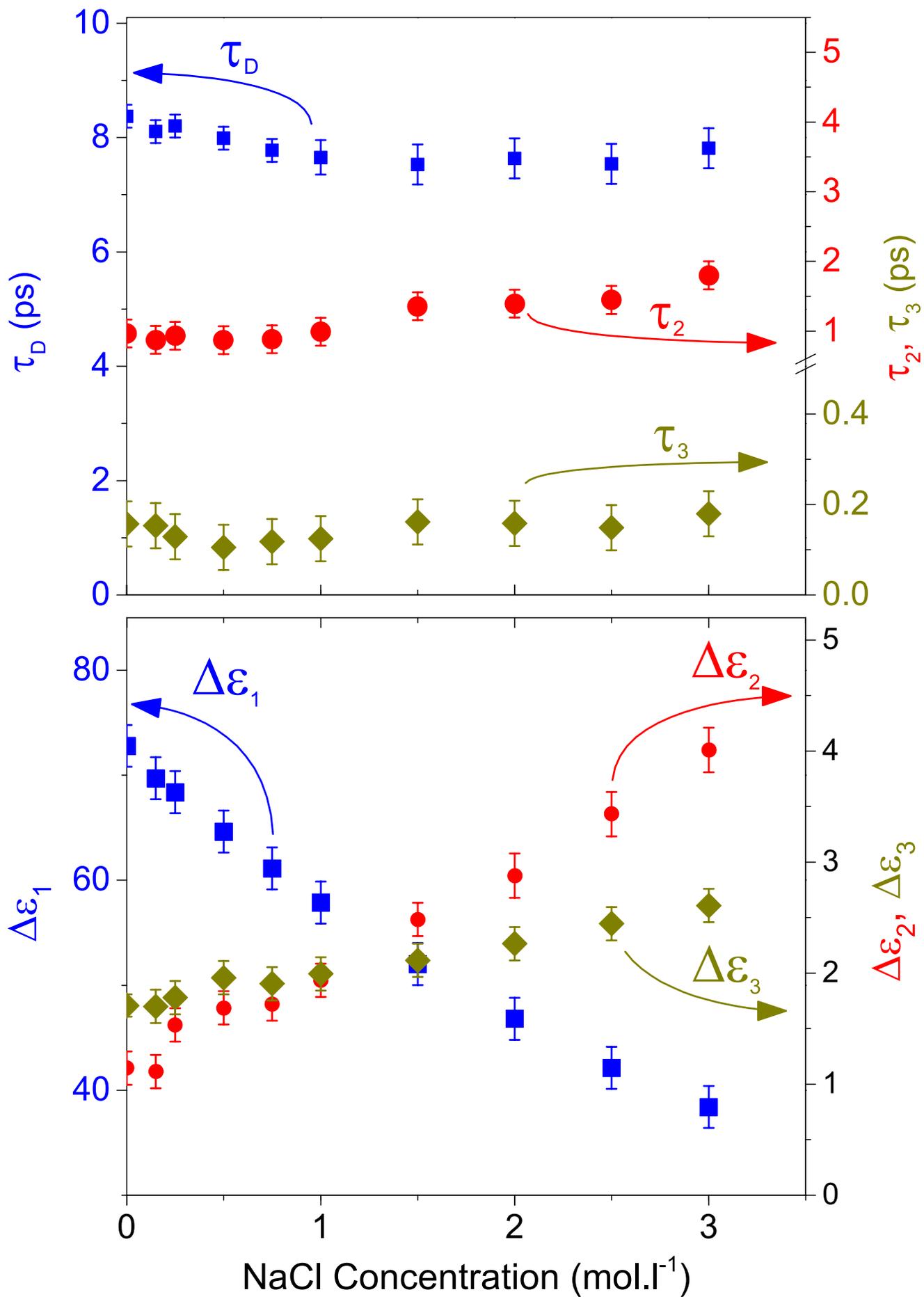

Figure 3

# High-precision gigahertz-to-terahertz spectroscopy of aqueous salt solutions as a probe of the femtosecond-to-picosecond dynamics of liquid water

N. Q. Vinh[1,2,4], Mark S. Sherwin[1,2], S. James Allen[1,2], D. K. George[4], A. J. Rahmani[4] and Kevin W. Plaxco[1,3]

[1]Institute for Terahertz Science and Technology, University of California, Santa Barbara Santa Barbara, CA 93106, USA

[2]Department of Physics, University of California, Santa Barbara, Santa Barbara, CA 93106, USA

[3]Department of Chemistry and Biochemistry, University of California, Santa Barbara

Santa Barbara, CA 93106, USA

[4]Department of Physics, Virginia Tech, Blacksburg, VA 24061, USA

Corresponding Author: vinh@vt.edu

**Supporting Information:**

*1. Comparison of log-log and linear fits*

Because the amplitude of the slowest relaxation phase is two orders of magnitude larger than those of the other, faster phases, it is over-weighted in simple, linear fits of our data, reducing the precision with which we can determine the properties of the (much smaller) faster phases. In this work we thus fit our data as a fit of $\log(\varepsilon^*)$ versus the logarithm of the dielectric response (i.e., we took the logarithm of both sides of eq. 2) in order to suppress the otherwise overwhelming contribution of the slowest phase to the global fit. This said, fits to eq. 2 (i.e., in its original, linear form) produce time constants and amplitudes within experimental uncertainty of those seen for the logarithmic fit, albeit with larger confidence intervals (table SI1). [Note: we also performed the fit that included a still slower mode at ~ 200 ps, which is suggested to arise due to ion-pair or ion-cloud relaxation[1]. Perhaps not surprisingly, given the vast difference in timescales (such a mode corresponds to a frequency of just 0.8 GHz) between this putative mode and the slowest mode we observe (8 ps), and given its rather small amplitude (at ~ 1.0 the reported amplitude of this mode is more than an order of magnitude lower than that of the 8-ps mode we observe), these fits do not differ significantly from fits to models lacking this putative mode.]

**Table SI1: a comparison of log-log and linear fits to equation 2 for water.**

| Parameter | Two-Debye log-log fit | Two-Debye linear fit | Three-Debye log-log fit | Three-Debye linear fit |
|---|---|---|---|---|
| $\varepsilon_1$ | 5.48 ± 0.10 | 5.76 ± 0.50 | 6.05 ± 0.10 | 5.95 ± 0.20 |
| $\varepsilon_2$ | | | 4.91 ± 0.10 | 4.60 ± 0.20 |
| $\varepsilon_\infty$ | 3.5 ± 0.3 | 3.7 ± 0.5 | 3.2 ± 0.3 | 2.7 ± 0.5 |
| $\tau_D$ | 8.23 ± 0.25 ps | 8.36 ± 0.25 | 8.37 ± 0.25 ps | 8.39 ± 0.25 ps |
| $\tau_2$ | 410 ± 150 fs | 420 ± 150 fs | 1.05 ± 0.15 ps | 0.90 ± 0.15 ps |
| $\tau_3$ | | | 178 ± 50 fs | 120 ± 70 fs |

## 2. Measurement of conductivity ($\sigma$)

We have measured the electrical conductivity of NaCl solutions using a portable EC/TDS/NaCl/Temperature Meter (HI98360N) from Hanna instruments. The instrument allows us to measure conductivity over the range 0 to 50 S/m with high accuracy.

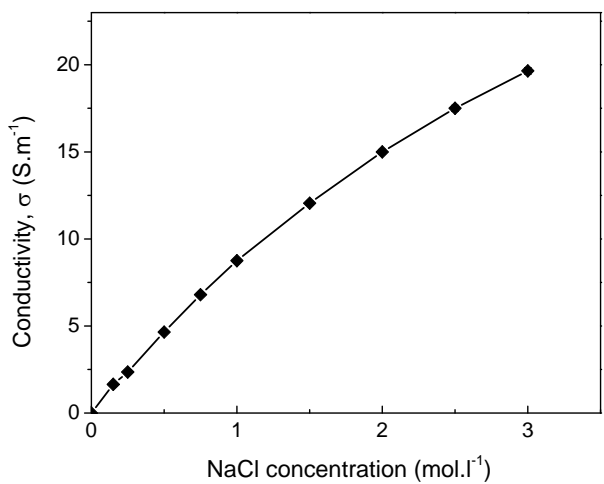

**Figure SI1:** The conductivity of NaCl solutions increases monotonically with increasing salt concentration.

## 3. Fit residuals

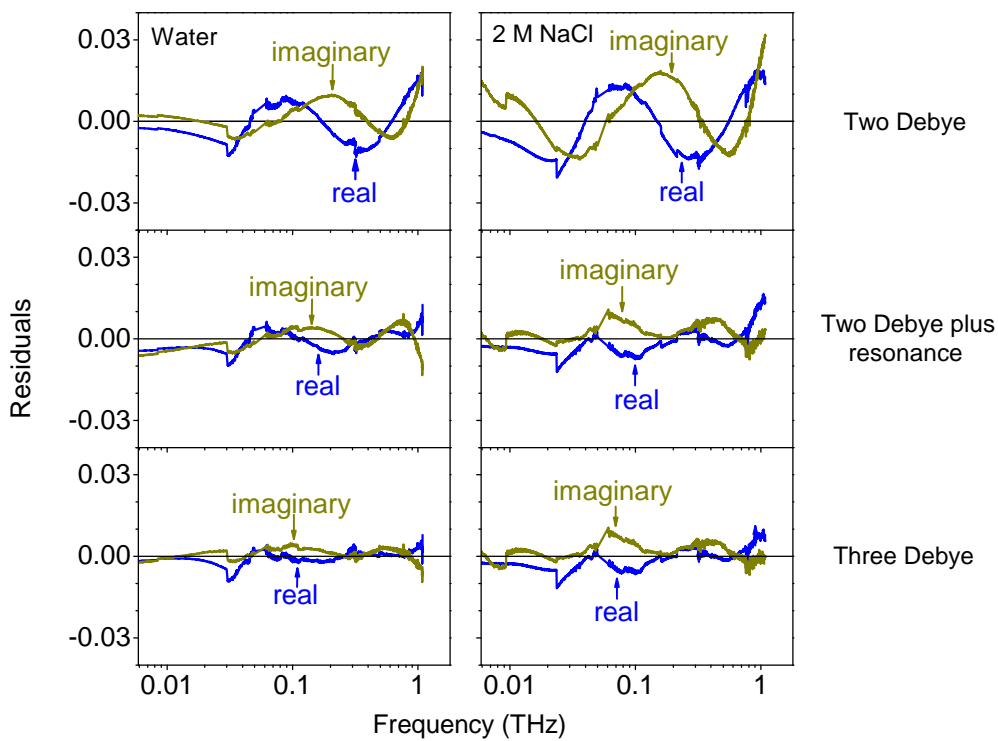

**Figure SI2**: Shown are the residuals for fits of (**top row**) the two-Debye model, (**middle row**) the two-Debye-plus-resonance model, and (**bottom row**) the three-Debye model for (**left**

**column**) water and (**right column**) 2 M NaCl. Even when taking into account the two additional parameters (relative to the two-Debye model), the three-Debye model achieves a statistically significantly improved fit (see discussion in text). While improved over the two-Debye model, the fit of the two-Debye-plus-resonance model is nevertheless poorer than that of the three-Debye model even though the latter contains one more fitted parameter.

**Table SI2:** A comparison of fit statistics for two-Debye, two-Debye-plus-resonance, and three-Debye models for water and 2 M NaCl. The three-Debye model provides a statistically significant improvement in the fits of both the real and the imaginary components relative to other two models.

**Water**

|  | Two-Debye model | Two-Debye-plus-resonance model | Three-Debye model |
|---|---|---|---|
| Number of floating parameters in model | 5 | 7 | 7 |
| Adjusted $R^2$ | 0.99979 | 0.99993 | 0.99997 |
| Reduced $\chi^2$ | $3.65\times10^{-5}$ | $1.30\times10^{-5}$ | $4.65\times10^{-6}$ |
| Root-mean-squared residuals: real component | 0.007 | 0.004 | 0.002 |
| Root-mean-squared residuals: imaginary component | 0.005 | 0.004 | 0.002 |
| Serial correlation of residuals ($R^2$): real component | 0.995 | 0.980 | 0.954 |
| Serial correlation of residuals ($R^2$): imaginary component | 0.992 | 0.985 | 0.945 |

**2 M NaCl**

|  | Two-Debye model | Two-Debye-plus-resonance model | Three-Debye model |
|---|---|---|---|
| Adjusted $R^2$ | 0.99909 | 0.99788 | 0.99990 |
| Reduced $\chi^2$ | $1.17\times10^{-4}$ | $2.72\times10^{-4}$ | $1.32\times10^{-5}$ |
| Root-mean-squared residuals: real part | 0.011 | 0.014 | 0.004 |
| Root-mean-squared residuals: imaginary component | 0.011 | 0.019 | 0.003 |
| Serial correlation of residuals ($R^2$): real component | 0.997 | 0.998 | 0.976 |
| Serial correlation of residuals ($R^2$): imaginary component | 0.996 | 0.998 | 0.946 |

*4. Two-Debye-plus-resonance model and salt solution data.*

We also fitted our salt solution data to a two-Debye-plus-resonance model ($n = 2$):

$$\varepsilon^* = \varepsilon_\infty + \Delta\varepsilon_u + \Delta\varepsilon_{sw} + \sum_{j=1}^{n}\frac{(\varepsilon_{j-1} - \varepsilon_j)}{1 + i2\pi\nu\tau_j} + i\frac{\sigma}{2\pi\nu\varepsilon_0} + \frac{A_{OSC}/(2\pi)^2}{\nu_{OSC}^2 - \nu^2 + i\nu(k_{OSC}/2\pi)} \quad \text{(SI1)}$$

For this fit we vary the parameters $\varepsilon_1$, $\varepsilon_\infty$, $\tau_D$, $\tau_2$, $A_{OSC}$, $\nu_{OSC}$ and $k_{OSC}$ but fix the electrical conductivity, $\sigma$, to our independently measured values (see Fig. SI1) and the static dielectric constant, $\varepsilon_s$, to previous literature values[2]. For 2 M NaCl we obtain best-fit values of $\varepsilon_1 = 6.76 \pm 0.20$, $\varepsilon_\infty = 3.9 \pm 0.3$, $\tau_D = 7.29 \pm 0.55$ ps, $\tau_2 = 0.42 \pm 0.15$ ps, $A_{OSC}/(2\pi)^2 = 6.14 \pm 0.50$ THz$^2$, $\nu_{OSC} = 1.1 \pm 0.1$ THz, and $k_{OSC} = 2.2 \pm 0.5$ THz. We find, however, that although the fit is reasonably good, the best-fit values of both $\tau_D$ and $\tau_2$ differ rather significantly with estimates from the prior literature (Table 1). Moreover, the fit statistics for the two-Debye-plus-resonance model (adjusted $R^2 = 0.99788$ and $\chi^2 = 2.74 \times 10^{-4}$) are significantly poorer than those for the three-Debye model (adjusted $R^2 = 0.99990$ and $\chi^2 = 1.32 \times 10^{-5}$), as are its fit residuals (root-mean-squared residuals: 0.014 for the real component and 0.019 for the imaginary component; Fig. SI2 and table SI2) and their serial correlations ($R^2$ of 0.998 for both the real and imaginary components). It thus appears that, as it does for pure water, the three Debye model accurately captures the femtosecond to picosecond relaxational dynamics of water even in the presence of very high concentrations of salt.

*5. The size the of hydration shell*

The salt dependence of the amplitude of the slowest (8 ps) relaxation provides a means of determining the number of water molecules, $N(c)$, so tightly bound to ions that they do not contribute to this relaxation process. This analysis has been performed for a large number of ionic solutions by Buchner *et al.*[2] and Kaatze *et al.*[3] using the relationship:

$$N(c_{el}) = \left(c_s - \frac{\Delta\varepsilon_{sol} + \Delta_{kd}\varepsilon(c)}{\Delta\varepsilon_{pure}} c_{pure}\right)/c_{el} \quad \text{(SI2)}$$

where $c_s$ is the concentration of water in the solution, $c_{pure} = 55.56$ M is the molar concentration of pure water, $\Delta\varepsilon_{sol} = \varepsilon(c_{el}) - \varepsilon_\infty(c_{el})$ is the dielectric strength of the solution at the NaCl concentration $c_{el}$, $\Delta\varepsilon_{pure}$ is the dielectric strength of pure water, and $\Delta_{kd}\varepsilon(c_{el})$ is the decrease of the relaxation strength of an ion due to the kinetic depolarization effect at the given NaCl concentration. This last term depends on the NaCl concentration via the relationship:

$$\Delta_{kd}\varepsilon(c_{el}) = \frac{2}{3}\frac{\varepsilon(0) - \varepsilon_\infty(c_{el})}{\varepsilon(0)}\frac{\tau_D(0)}{\varepsilon_0}\sigma(c_{el}) \quad \text{(SI3)}$$

where $\sigma$ is the electrical conductivity of the solution (Fig. SI1). As noted in the text, the results we obtain suggest that $4.8 \pm 0.5$ water molecules are bound per equivalent of electrolyte, a value quite close to previous reports[2,4].

*6. The gigahertz-to-terahertz absorption and refractive index of pure liquid water and aqueous salt solutions.*

Download at [ftp://ftp.aip.org/epaps/journ_chem_phys/E-JCPSA6-142-045516/SI.pdf](ftp://ftp.aip.org/epaps/journ_chem_phys/E-JCPSA6-142-045516/SI.pdf)

**References:**

1. A. Eiberweiser and R. Buchner, J Mol Liq **176**, 52-59 (2012).
2. R. Buchner, G. T. Hefter and P. M. May, J Phys Chem A **103**, 1-9 (1999).
3. U. Kaatze, J Solution Chem **26**, 1049-1112 (1997).
4. K. Nortemann, J. Hilland and U. Kaatze, J Phys Chem A **101**, 6864-6869 (1997).